\documentclass[manuscript,screen]{acmart}

\AtBeginDocument{%
  }

\setcopyright{acmlicensed}
\copyrightyear{2018}
\acmYear{2018}
\acmDOI{XXXXXXX.XXXXXXX}

\acmConference[Conference acronym 'XX]{Make sure to enter the correct
  conference title from your rights confirmation emai}{June 03--05,
  2018}{Woodstock, NY}
\acmISBN{978-1-4503-XXXX-X/18/06}

\usepackage{multirow}
\usepackage{makecell}
\usepackage{tabularx}
\usepackage[T1]{fontenc}
\usepackage{array}
\usepackage{listings}
\usepackage{xspace} 
\usepackage{subfigure}
\usepackage{textcomp}
\usepackage{microtype} 
\usepackage{calc}
\usepackage{pifont}
\usepackage{tcolorbox}

\usepackage{adjustbox}
\usepackage{amsmath,amssymb,amsfonts}
\usepackage{algorithmic}
\usepackage{graphicx}
\usepackage{textcomp}
\usepackage{xcolor}
\usepackage{multirow}
\usepackage{mdframed}
\usepackage{balance}
\usepackage{tcolorbox}
\usepackage{makecell}
\usepackage{threeparttable}
\usepackage{colortbl}
\usepackage{subfigure}
\usepackage{hhline}
\usepackage{pifont}
\usepackage{listings}
\usepackage{enumitem}

\newcommand{\parabf}[1]{\noindent\textbf{#1}}

\newcommand{\Comment}[1]{}

\newcommand{\red}[1]{\textcolor{red}{+#1}}
\newcommand{\blue}[1]{\textcolor{green}{-#1}}

\newcommand{\rev}[1]{\textcolor{black}{#1}}

\newcommand{\app}{IntDiagSolver}

\definecolor{ggray}{HTML}{eff0f0}
\definecolor{gggray}{HTML}{E8E8E8}
\definecolor{ggggray}{HTML}{BEBEBE}
\definecolor{myblue}{RGB}{255,255,255}
\definecolor{myyellow}{HTML}{FFF2CC}
\definecolor{myfinding}{HTML}{E7F1FA}
\definecolor{myanswer}{HTML}{FDDECE}

\newcommand{\ie}{\textit{i.e.,}}
\newcommand{\eg}{\textit{e.g.,}}
\newcommand{\bug}{crash bugs}
\newcommand{\codebug}{code-related crash bugs}
\newcommand{\envbug}{environment-related crash bugs}
\newcommand{\crashinfo}{crash information}

\usepackage[utf8]{inputenc}
\usepackage[english]{babel}

\newcounter{finding}

\begin{document}

\title{Exploring Large Language Models in Resolving Environment-Related Crash Bugs: Localizing and Repairing}

\author{Xueying Du}
\email{xueyingdu21m.fudan.edu.cn}
\orcid{1234-5678-9012}
\affiliation{%
  \institution{Department of Computer Science, Fudan University}
  \city{Shanghai}
  \country{China}
}

\author{Mingwei Liu}
\email{liumw26@mail.sysu.edu.cn}
\affiliation{%
  \institution{School of Software Engineering, Sun Yat-sen University}
  \city{Zhuhai}
  \country{China}
}

\author{Hanlin Wang}
\email{wanghanlin23@m.fudan.edu.cn}
\affiliation{%
  \institution{Department of Computer Science, Fudan University}
  \city{Shanghai}
  \country{China}
}

\author{Juntao Li}
\email{22210240197@m.fudan.edu.cn}
\affiliation{%
  \institution{Department of Computer Science, Fudan University}
  \city{Shanghai}
  \country{China}
}

\author{Xin Peng}
\email{pengxin@fudan.edu.cn}
\affiliation{%
  \institution{Department of Computer Science, Fudan University}
  \city{Shanghai}
  \country{China}
}

\author{Yiling Lou}
\email{yilinglou@fudan.edu.cn}
\affiliation{%
  \institution{Department of Computer Science, Fudan University}
  \city{Shanghai}
  \country{China}
}





\begin{abstract}
 Software crash bugs cause unexpected program behaviors or even abrupt termination, thus demanding immediate resolution. However, resolving crash bugs can be challenging due to their complex root causes, which can originate from issues in the source code or external factors like third-party library dependencies. Large language models (LLMs) have shown promise in software engineering tasks, leveraging their extensive pre-training on text and code corpora. However, existing research predominantly focuses on the capability of LLMs to localize and repair code-related crash bugs, leaving their effectiveness in resolving environment-related crash bugs in real-world software unexplored.

To fill this gap, we conducted the first comprehensive study to assess the capability of LLMs in
resolving real-world environment-related crash bugs, \rev{using a newly constructed dataset of 100 representative crash bugs.} We first systematically compare LLMs’ performance in resolving code-related and environment-related crash bugs with varying levels of crash contextual information. Our findings reveal that the LLM performs better in resolving code-related crash bugs compared to environment-related ones. Specifically, localization is the primary challenge for resolving code-related crashes, while repair poses a greater challenge for environment-related crashes. Furthermore, we investigate the impact of \rev{different prompt strategies on improving the resolution of environment-related crash bugs, incorporating different prompt templates and multi-round interactions. Building on this, we further explore an advanced active inquiry prompting strategy, which leverages the self-planning capabilities of LLMs to conduct systematic and continuous questioning aimed at identifying potential environmental factors that contribute to crashes. Based on these explorations, we propose \app{}, an interactive methodology designed to enable precise crash bug resolution through ongoing engagement with LLMs. Extensive evaluations of \app{} on a dataset of 41 crash bugs across multiple LLMs (including GPT-3.5, GPT-4, Claude, CodeLlama, DeepSeek-R1, and Qwen-3-Coder) demonstrate consistent improvements in resolution accuracy, with substantial enhancements ranging from 9.1\% to 43.3\% in localization and 9.1\% to 53.3\% in repair. Furthermore, the strong performance of \app{} on the latest expanded multilingual dataset of 42 crash bugs highlights its strong generalizability and effectiveness on previously unseen data.}

\end{abstract}






\maketitle

\section{Introduction}

Crash bugs pose significant challenges, potentially leading to program termination and disrupting user operations~\cite{icsm11crashresolve2, ase15crashresolve}. These failures typically originate from two distinct categories: code-related defects (\ie{} bugs in the source code, such as null pointer dereferences) and external environmental factors (\eg{} improper OS versions, or missing third-party dependencies)~\cite{icse21crasolver}. 

Although extensive efforts have been made in automatically localizing and repairing code-related crash bugs\cite{cao2020fl, ase2021fl, icse22fl, msr21apr, arXiv23chatgptrepair, arXiv23ConversationalProgramRepair}, addressing environment-related crash bugs remains challenging. This complexity arises from the multifaceted nature of environmental dependencies, including system configuration discrepancies, version incompatibilities, permission constraints, and hardware-specific requirements. Moreover, the numerous environmental factors involved within a software system make achieving an end-to-end solution more challenging. Existing techniques often rely on adopting similar solutions from online forums like Stack Overflow (SO)~\cite{icse21crasolver,fse20masetro, icst22masetroextend}, which lack generalizability, particularly for previously undocumented issues. Our preliminary analysis of 100 crash-related Stack Overflow threads reveals that 68\% of root causes stem from environmental factors, underscoring the prevalence and practical significance of this issue.

Recent advances in large language models (LLMs) demonstrate remarkable capabilities in code comprehension~\cite{wang2020codesummarization,23codesummarization} and generation tasks~\cite{poesia2022codegeneration,zeng2022codegeneration, du2023classeval}. While some existing researches have explored LLMs' potential in bug fixing\cite{arXiv21chatgptfixbug, arXiv23ConversationalProgramRepair}, these efforts predominantly focus on code-related crash bugs. The effectiveness of LLMs in resolving environment-related crash bugs in real-world software remains unexplored.

To address these knowledge gaps, we conducted the first comprehensive study to assess the capability of LLMs in resolving real-world environment-related crash bugs. Our investigation focused on evaluating LLMs' effectiveness in both localizing and repairing environment-related crash bugs using various prompts and interaction methods. Specifically, we: (i) systematically compared the performance of LLMs in resolving both code-related and environment-related crash bugs under different levels of contextual information, \rev{(ii) investigated how various prompt strategies on improving the resolution of \envbug{}, incorporating different prompt templates and multi-round interactions,  and (iii) further explored the effectiveness of a novel active inquiry prompting strategy that leverages self-planning to enhance multi-round prompting.} We aim to answer the following two research questions:

\begin{itemize}
\item \textbf{RQ1 (Comparative Effectiveness)}: How do LLMs perform in localizing and repairing code-related versus environment-related crash bugs with varying levels of contextual information?

\item \rev{\textbf{RQ2 (Enhancing Resolution with Prompt Engineering)}: How do existing prompt strategies, as well as our proposed advanced prompt strategy, improve the effectiveness of LLMs in resolving environment-related crash bugs?}

\begin{itemize}
\item \textbf{RQ2.a (Influence of Prompt Templates)}: How do different prompt templates affect LLMs' ability to resolve crash bugs?

\item \rev{\textbf{RQ2.b (Impact of Multi-Round Prompts)}: What is the impact of employing multi-round prompts on the effectiveness of LLMs in resolving crash bugs?}

\item \rev{\textbf{RQ2.c (Impact of Active Inquiry Prompt with Self-Planning)}: What is the impact of employing the advanced active inquiry prompt on the effectiveness of LLMs in resolving crash bugs?}

\end{itemize}
\end{itemize}

Our study of 100 real-world crash bugs reveals key findings.
Firstly, the LLM demonstrates more excellent proficiency in resolving \textit{\codebug{}} than \textit{\envbug{}}. Secondly, we identify fault localization as the primary challenge in LLMs' resolution of \codebug{}, while repair poses a greater challenge for \envbug{}. Thirdly, continuous interaction with LLM proves to be effective in acquiring relevant knowledge and enhancing the effectiveness of environment-related crash bug resolution. Moreover, by stimulating LLMs' proactive questioning ability for self-guided planning, we observe a significant improvement in resolution as the LLM navigates the resolution process through step-by-step localization.

Based on our findings, we introduce a methodology \app{} to enhance the interaction with LLMs for more accurate resolution of crash bugs. While our primary focus is on \envbug{}, we ensure the applicability of our approach to code-related crash bugs by proposing distinct prompt templates highlighting different aspects of crash contexts for both types of crash bugs respectively, coupled with diverse multi-round interaction strategies. Additionally, we devise a strategy enabling LLMs to guide the repair process, which is particularly beneficial for users with limited crash bug knowledge.
We then evaluate our proposed methodology addressing the following research questions:

\begin{itemize}
    \item \rev{\textbf{RQ3 (Effectiveness of \app{})}: How does \app{} enhance the ability of LLMs to effectively resolve crash bugs?}
    \item \rev{\textbf{RQ4 (Generalizability of \app{})}: How well does \app{} generalize to different scenarios?}
    \begin{itemize}
        \item \rev{\textbf{RQ4.a (Generalizability across LLMs)}: To what extent can the effectiveness of \app{} in resolving crash bugs be generalized across different LLMs?}
        \item \rev{\textbf{RQ4.b (Generalizability across Programming Languages)}: To what extent can the effectiveness of \app{} in resolving crash bugs be generalized across different programming languages?}
    \end{itemize}
\end{itemize}

The results demonstrate that \app{} consistently improves the accuracy of resolving both code-related and environment-related crash bugs \rev{across six different LLMs, with substantial enhancements ranging from 9.1\% to 43.3\% in localization and 9.1\% to 53.3\% in repair.} Notably, there is a remarkable improvement from 0/30 to 16/30 in the number of successfully repaired environment-related crash bugs on GPT-3.5. \rev{Furthermore, results on the latest expanded multilingual environment-related crash bug dataset further highlights the strong generalizability and effectiveness of \app{} on previously unseen data, with localization improvements of 12.0\% to 16.6\% and repair improvements of 19.0\% to 21.4\% across two advanced LLMs.}

The contributions of this work are summarized as follows:

\begin{itemize}
\item \textit{First Comprehensive Study}: Represents the first comprehensive investigation into LLMs' ability to localize and repair \envbug{}. Our benchmark and results are available at online replication package~\cite{replication_package}.
\item  \textit{Insightful Findings}: Delves into the strengths and limitations of crash bug resolution using LLMs, providing a comparative analysis of LLM performance in code-related and \envbug{} localization and repair. Furthermore, we explore the impact of different prompt enhancement techniques and interaction methods on resolution outcomes. 
\item  \textit{Innovative Methodology}: Introduces the novel methodology \app{} that enhances the interaction with LLMs for more accurate crash bug resolution.
\end{itemize}

\section{RQ1: Comparative Effectiveness} 

In this RQ, we perform a comparative analysis of the LLM's ability to localize and repair code-related and environment-related crash bugs, considering varying levels of crash contextual information.

\rev{\textbf{Code-related crash bugs} refer to those caused by defects within the software’s source code. Repairing such bugs typically involves modifying the source code itself. In contrast, \textbf{environment-related crash bugs} arise from issues or incompatibilities in external factors required for the software’s execution. These may include the operating system version or configuration, dependency library versions, hardware resources, network status, or the availability and configuration of third-party services. Typical resolutions for environment-related bugs involve adjusting the relevant environmental configurations, such as modifying configuration files or updating dependencies, without the need to change the source code.}

\subsection{Benchmark}\label{sec:benchmark}

Previous researches~\cite{arXiv23ConversationalProgramRepair, arXiv23AutomaticBugFixing, prenner2021automatic} exploring the potential of LLMs in bug resolution typically utilize benchmarks such as QuixBugs~\cite{QuixBug}, Defects4j~\cite{issta2014Defects4J}, ManySStuBs4J~\cite{karampatsis2020often}, and UnifiedBugDataset~\cite{ferenc2018public}. However, these benchmarks primarily focus on bugs originating from faulty source code, neglecting those arising from environmental issues, which are prevalent and sometimes more challenging for developers~\cite{Zhao23Knowledge, wang2023automatically}. Furthermore, the widely used QuixBugs benchmark, which includes only 40 bugs in relatively simple scenarios, such as basic algorithms, fails to capture the complexity of real-world crash bugs.

To comprehensively compare the effectiveness of LLMs in addressing code-related and environment-related crash bugs in real-world applications, we create a new benchmark derived from SO threads, incorporating both code-related and environment-related issues. SO provides rich contextual information on buggy code's purposes and dependencies.

\subsubsection{Benchmark Format}
\label{sec:benchmark_format}

\begin{figure}[htb] 
    \centering   
    \vspace{-3mm}
    \subfigure[Code-related Bug] 
    {  
        \begin{minipage}[t]{0.46\linewidth}
        \centering  
        \includegraphics[width=\linewidth]{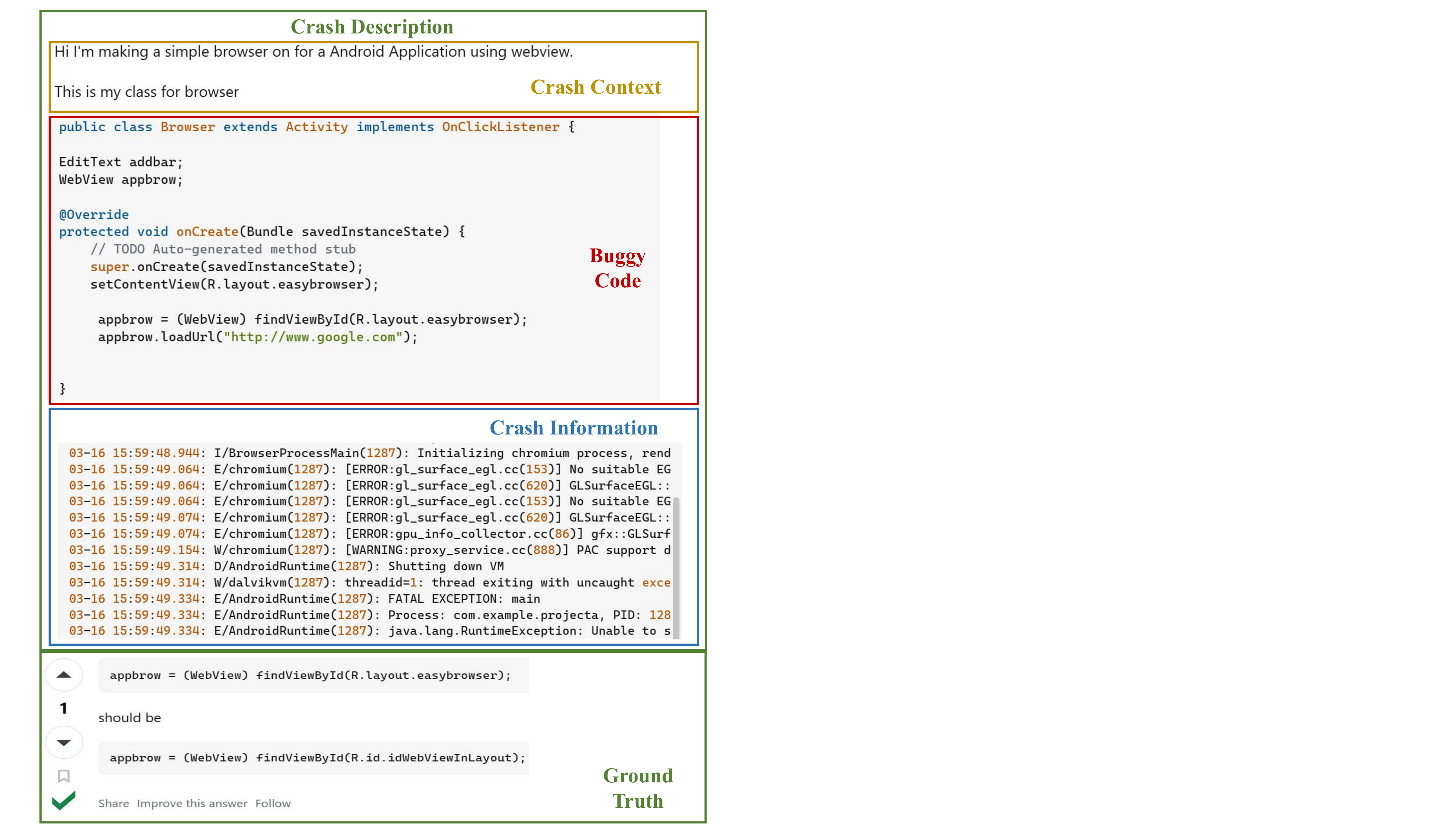}  
        \label{fig:crash_bug}
        \end{minipage}
    }     
    \subfigure[Environment-related Bug] 
    {    
    \begin{minipage}[t]{0.485\linewidth}
    \centering 
    \includegraphics[width=\linewidth]{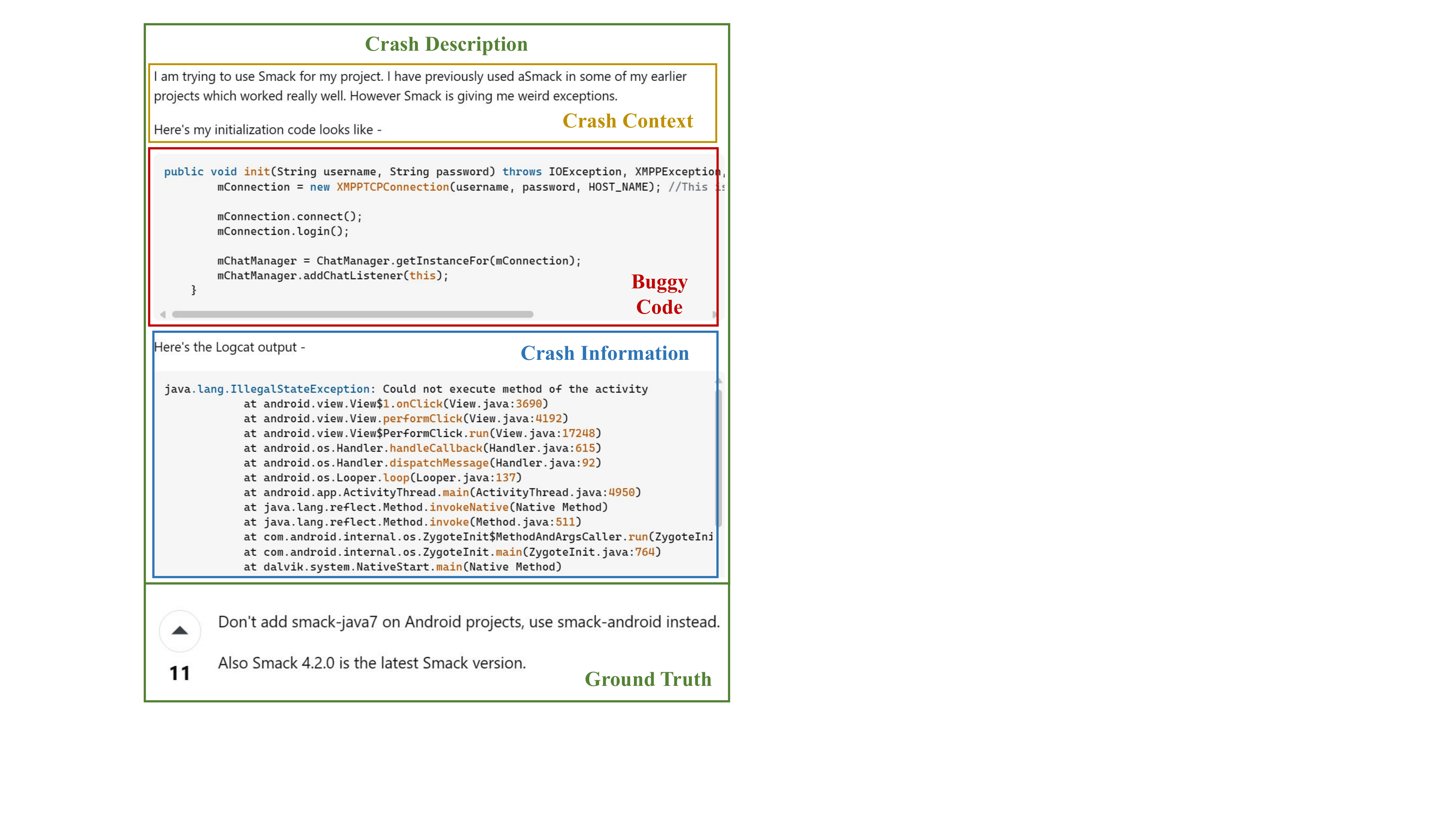}    
    \label{fig:env_crash_bug}
    \end{minipage}
    }   
    \vspace{-2mm}
    \caption{Examples of Crash Bugs} 
    \label{example_of_crash_bug}   
    \vspace{-3mm}
\end{figure}

We extracted crash descriptions (Cra-Des) for each crash bug from the question sections of SO threads. 
These descriptions
facilitate a comprehensive analysis of LLMs' effectiveness in localizing and repairing crashes within various information contexts. Specifically, crash descriptions consist three primary parts of information:

\begin{itemize}
    \item \textbf{Buggy Code (B-Code) }: This snippet of code triggers the crash and often contains syntax or semantic errors that lead to unexpected program behavior.

    \item \textbf{Crash Information (Cra-Info) }: This segment includes the exception type, error message, and crash trace. Exception type identifies the error, \eg{} divide-by-zero. The error message provides a human-readable description of the error. The crash trace offers a stack trace of the program execution at the time of the crash, assisting in locating the error.

    \item \textbf{Crash Context (Cra-Cont) }: This section provides a natural language description of the broader crash context, including symptoms, potential causes, relevant inputs, the buggy code's purpose, configurations, or dependencies possibly contributing to the error. Such details are crucial for replicating the crash and grasping its underlying cause. 
\end{itemize}

We obtained the ground truth for localizing and repairing each crash bug from the accepted answers in the corresponding SO thread. Figure~\ref{fig:crash_bug} illustrates examples of both code-related and environment-related crash bugs from our benchmark, specifically extracted from SO threads \#22442280 and \#30322026.

\subsubsection{Benchmark Construction Procedure}

We automated the construction of a pool of SO threads enriched with detailed crash descriptions, from which we selected 100 samples possessing high-quality ground truths to serve as our dataset. In alignment with previous research~\cite{ase15crashresolve, icse21crasolver}, we focus on crash bugs in the Java program due to their prevalence~\cite{fse20masetro}.

\textbf{Establishing Data Pool.}
To establish a benchmark encompassing sufficient contextual information for both code-related and environment-related crash localization and repair, we selected high-quality threads from the SO data dumps~\cite{so_dump} based on the following inclusion criteria: 
(1) Thread titles or tags containing the keyword ``Java'', (2) Thread titles or tags containing the keywords ``exception'' or ``error'',  (3) Threads with an accepted answer, (4) Threads with at least one positive vote for their question, (5) Threads with concrete error symptoms (\ie{} code or stack trace) to ensure sufficient context for understanding the crash bug, and (6) Threads containing at least one common Java exception type (\eg{} \textit{NullPointerException}). 
Furthermore, to collect a comprehensive list of common Java exception types for Criterion-6, we systematically parsed 35,773 Java libraries from Maven Central~\cite{mavencenter} based on the Libraries.io dataset~\cite{libariesiodata} and JDK 1.8~\cite{java_api}. 
In this way, we collected 67,248 SO threads related to crash bugs as the pool.

\textbf{High-Quality Sample Selection.}
We initially randomly sampled 100 threads from the curated pool, \rev{and then two authors independently assessed the accepted answers in each thread, which are the answers verified for correctness by the original question askers. Following prior work\cite{accanswer1, accanswer2, accanswer3, accanswer4}, we treated these accepted answers as the ground truth solutions for the corresponding crash bugs.} Only threads with ground truth solutions consistently rated as high quality by both assessors were included in the final benchmark. High-quality solutions were defined as those providing detailed code modifications or problem-solving steps, in contrast to merely offering vague troubleshooting tips or general fixing directions without specific guidance. \rev{Additionally, we filtered out cases that could not be unambiguously categorized as either code-related or environment-related.} 
We continued sampling and assessment until we had 50 qualified SO threads for both code-related and environment-related types, \ie{} 100 crash bugs in total. \rev{Since the cases were randomly selected, the difficulty levels of the code-related and environment-related cases are expected to follow a natural probability distribution. Table~\ref{table:dataset} summarizes the statistics of our dataset.}

\begin{table*}[htb]
	\centering
	\vspace{-2mm}
    \caption{
    \rev{Dataset Statistics. The ``Buggy code lines'' column shows the average number of buggy code lines per instance, while the ``Crash trace length'' and ``Crash description length'' indicate the average token counts of the crash trace and the natural language description in the corresponding thread, respectively.}}
    \label{table:dataset}
    \footnotesize
    \begin{adjustbox}{width=0.8\columnwidth}
    \begin{tabular}{|m{3cm}<{\centering}|m{3cm}<{\centering}|m{3cm}<{\centering}|m{3cm}<{\centering}|} \hline
    \textbf{Type} & \textbf{Buggy Code Lines} & \textbf{Crash Trace Length} & \textbf{Crash Description Length} \\ \hline
    Code-related & 47.24 & 783 & 418 \\ 
    Environment-related & 34.57 & 2773 & 406 \\ \hline
    \end{tabular}
    \end{adjustbox}
    \vspace{-2mm}
\end{table*}

\subsection{Setup and Metrics}
\label{rq1:metrics}

We manually reviewed LLM's responses and conducted separate qualitative analyses to assess its localization and repair capabilities. We opted for the gpt-3.5-turbo-1106 model for this RQ. This choice was driven by its more rapid response and cost-effectiveness~\cite{sun2023gpt}.

\textbf{Correctness of Localization and Repair.}
\textit{Code-related Crash Bugs.} In the localization evaluation, a solution was deemed correct if the line of code it identified for modification matched the faulty code line identified in the benchmark. In the repair evaluation, the LLM was required to provide an accurate code patch that matched the one in the benchmark to be considered correct. 
\textit{Environment-related Crash Bugs.} In the localization evaluation, the LLM's localization was considered correct if it identified the precise cause of the exception (\eg{} a library version problem or a missing dependency problem). In the repair evaluation, a specific repair plan was required (\eg{} upgrading or downgrading to a specific library version).

The evaluation was conducted independently by two authors. 
In cases of disagreement, a third evaluation was carried out using the same prompt to determine the correctness of the response, following a majority-win strategy.
Consequently, the Kappa coefficient~\cite{kappa} for code-related crash bugs localization and repair was 0.89 and 0.92 respectively; for both localization and repair in \envbug{} was 0.97, indicating almost perfect agreement.
It's worth noting that only a small percentage of queries (5\%, 5 out of 100) required this third evaluation to reach a consensus.

\textbf{Solution Number.} Typically, the LLM produces a single accurate solution with high confidence in its diagnosis; With reduced confidence, it may propose multiple potential solutions. 
Consequently, we recorded the number of solutions the LLM provided for each crash bug and manually verified the correctness of localization and repair for each solution.

\textbf{Accuracy \& Success.} We calculated the accuracy rates for localization and repair, reflecting the proportion of correct solutions for localization/repair to the total solutions offered. The formula for calculating accuracy is as follows:
\vspace{-2mm}
\begin{equation}
Accuracy  = 
\frac{1}{N} \left(\sum_{i=1}^{N} \frac{Correct Sol. Number_{i}}{Total Sol. Number_{i}}\right) \times 100 \%
\end{equation}

A crash bug is deemed successfully localized/repaired if any one of the solutions within a response is identified as correct. 

\subsection{Resolving Code-related Crash Bugs}
\label{sec:rq1_code}

\subsubsection{Design}
\label{sec:rq1_design}
We first assess the LLM's ability to localize and repair 50 code-related crash bugs in our benchmark, 
utilizing the aforementioned three types of crash description information (\ie{} Buggy Code, Crash Information, and Crash Context). 

\parabf{Prompt Design.}
A series of prompt templates (Basic-Prompt-1 to Basic-Prompt-4) were designed for interacting with the LLM, covering different levels of granularity of information. 

\begin{mdframed}
[linecolor=myblue!50,linewidth=2pt,roundcorner=10pt,backgroundcolor=myyellow!20]
\small

\hspace{3.5mm}\textbf{Basic-Prompt-1:} This is my code: [B-CODE] I'm getting an exception, how do I fix it?

\textbf{Basic-Prompt-2:} This is my code: [B-CODE] I'm getting [CRA-INFO], how do I fix it?

\textbf{Basic-Prompt-3:} [CRA-CONT] This is my code: [B-CODE] I'm getting [CRA-INFO], how do I fix it?

\textbf{Basic-Prompt-4:} [CRA-DES] This is my error lines: [LOC] how to fix it?

\end{mdframed}

\parabf{Interaction Procedure.}
For each code-related crash bug in our benchmark, we conduct experiments with Basic-Prompt-1 to 3 to evaluate the LLM's effectiveness with different levels of provided information. Basic-Prompt-1 includes only the buggy code, Basic-Prompt-2 adds the buggy code along with additional \crashinfo{}, and Basic-Prompt-3 further incorporates crash context. For crash bugs that could not be correctly resolved with any of the aforementioned prompts, Basic-Prompt-4 was employed. This prompt specifically pinpointed the \textit{localization result in the buggy code lines (LOC)}, identifying the exact lines requiring modification. This simplified the task for the LLM, allowing it to focus solely on repairing the crash bug without localization, thereby enabling an examination of the model's repair capabilities in isolation.

\subsubsection{Results}
\label{subsec:rq1_result}

\begin{table*}[htb]
	\centering
	\vspace{-2mm}
    \caption{
    Experimental Results on Crash Bug Resolution. The ``Information'' column contains input data, while the ``Localization'' and ``Repair'' columns show the number of successfully localized and repaired crash bugs. ``Loc. with Multi.'' indicates cases where the LLM localized the bug but offered multiple solutions.}
    \label{table:rq1_result}
    \footnotesize
    \vspace{-2mm}
    \begin{adjustbox}{width=1\columnwidth}
    \begin{tabular}{|m{1.5cm}<{\centering}|m{1.8cm}<{\centering}|m{3cm}<{\centering}|m{1.5cm}<{\centering}|m{1.5cm}<{\centering}|m{1.2cm}<{\centering}|m{2cm}<{\centering}|m{2cm}<{\centering}|} \hline
    \textbf{Type} & \textbf{Prompt} & \textbf{Information} & \textbf{Localization} & \textbf{Repair} & \textbf{Loc. with Multi.} & \textbf{Localization Accuracy} & \textbf{Repair Accuracy}\\  \hline
    \multirow{4}{*}{Code related} & Basic-Prompt-1 & B-Code & 20/50 & 20/50 & 2 & 0.380 & 0.380 \\ \cline{2-8}
    & Basic-Prompt-2 & B-Code + Cra-Info & 30/50 (\red{10}) & 28/50 (\red{8}) & 2 & 0.580 (\red{52.6\%}) & 0.540 (\red{42.1\%})\\ \cline{2-8}
    & Basic-Prompt-3 & B-Code + Cra-Info + Cra-Cont & 42/50 (\red{22}) & 39/50 (\red{19}) & 3 & 0.807 (\red{112.3\%}) & 0.747 (\red{96.6\%}) \\ \cline{2-8}
    & Basic-Prompt-4 & B-Code + Cra-Info + Cra-Cont + LOC &46/50 (\red{26}) & 43/50 (\red{23}) & 3 & 0.887 (\red{133.3\%}) &  0.827 (\red{117.5\%}) \\ \hline \hline 
    \multirow{2}{*}{Env. related} & Basic-Prompt-2 & B-Code + Cra-Info & 7/50 & 5/50 & 5 & 0.063 & 0.047 \\ \cline{2-8}
    & Basic Prompt 3 & B-Code + Cra-Info + Cra-Cont & 32/50 (\red{25}) & 20/50 (\red{15}) & 21 & 0.341 (\red{441.3\%}) & 0.193 (\red{310.7\%}) \\ \hline 
    \end{tabular}
    \end{adjustbox}
    \vspace{-1mm}
\end{table*}

Table \ref{table:rq1_result} illustrates the experimental results of the LLM's ability to resolve code-related crash bugs. The multifaceted analysis yields the following main conclusions (Finding 1-4). 

\parabf{Response Analysis.}
We collected the responses of the LLM for different prompts and manually analyzed the contents of the responses. For code-related crash bugs, the LLM's responses can encompass several aspects, including confirming code functionality, explaining error messages, identifying potential causes, offering textual fixes, presenting repaired code, discussing alternatives, and providing warnings.

\parabf{Overall Resolution Capability.} The experimental results presented in Table \ref{table:rq1_result} indicate that the LLM has significant potential in fixing \codebug{}. Specifically, when provided with full crash description information, the LLM was able to correctly localize 42 out of 50 crash bugs (84.0\%) and repaired 39 of them (78.0\%). Furthermore, the LLM provided a unique and clear answer for the majority of repaired crash bugs, with only 7.9\% (3 out of 38) repaired crash bugs providing multiple possible solutions.

\vspace{-3mm}
\begin{tcolorbox}[colback=myfinding!50, colframe=white, width=\linewidth, arc=3mm, boxrule=0.5mm, left=2mm, right=2mm, top=2mm, bottom=2mm, boxsep=0mm]
\textbf{Finding 1}: The LLM excels in resolving code-related crash bugs, displaying proficiency in bug localization and repair.
\end{tcolorbox}
\vspace{-3mm}

\parabf{Contextual Information Impact.} The LLM demonstrates the ability to resolve code-related bugs using only the information available within the buggy code. As shown in Table \ref{table:rq1_result}, the LLM successfully localized and repaired 40\% (20 out of 50) of crash bugs when provided with only the buggy code using Basic-Prompt-1.  This indicates that the LLM can effectively understand and resolve code-related bugs to a certain extent, relying solely on the information within the buggy code.

Moreover, Table \ref{table:rq1_result} illustrates that providing more detailed crash description information significantly enhanced the LLM's performance in localizing and repairing crash bugs. \textbf{Notably, both the number of localized and repaired bugs increased, indicating that a more effective crash description improved the LLM's bug localization accuracy.}

The significance of exception type and error message in crash localization is highlighted by the responses of the LLM and the analysis of two specific cases: \bug{} 68199510 \cite{68199510} and 65084069 \cite{65084069} (the SO thread IDs of the crash bugs). In both cases, the LLM provided environment-specific solutions, leading to misdiagnosis. This is because the exception types \textit{NoClassDefFoundError} and \textit{NoSuchMethodError} are commonly associated with environmental issues.
Our investigation further reveals that complete stack traces in \codebug{} play a relatively minor role compared to the exception type information. This finding was validated through additional experiments on specific crash cases. In crash bug 22928450 \cite{22928450}, the LLM successfully generated the correct repair solution when provided with only the buggy code, crash context, and exception type. However, when the exception type information was replaced with a stack trace that lacked explicit exception type details, the LLM failed to accurately localize the crash bug.
Therefore, \textbf{to improve repair effectiveness, it is beneficial to explicitly emphasize exception type information in natural language within the crash description.}

Comprehensive crash context information also enhances the LLM's ability to localize issues, encompassing various aspects such as the intended functionality of the program, detailed symptoms of the crash (e.g., specific conditions under which the crash occurs versus when the program operates normally, user-generated test cases, and previous unsuccessful repair attempts). 
Furthermore, it is worth noting that we observed no cases where a crash bug successfully resolved with limited information became unresolved when richer context was introduced. This indicates that providing more detailed context consistently improve bug resolution effectiveness.

\vspace{-3mm}
\begin{tcolorbox}[colback=myfinding!50, colframe=white, width=\linewidth, arc=3mm, boxrule=0.5mm, left=2mm, right=2mm, top=2mm, bottom=2mm, boxsep=0mm]
\textbf{Finding 2}: 
Enhanced crash description information significantly improves the LLM's ability to localize code-related bugs. Exception type and error message are crucial for precise localization, and additional context in natural language descriptions further enhances its performance.

\end{tcolorbox}
\vspace{-3mm}

\parabf{Localization vs. Repair.} 
Among the queries complete crash descriptions, only 7.1\% (3 out of 42) of correctly localized bugs remained unresolved. Moreover, in cases where the LLM initially failed to locate the crash bug, providing the actual buggy code line enabled successful repair in 4 out of 7 instances. This indicate that the LLM's resolving bottleneck mainly lies in localization. Once the LLM correctly localize the crash bugs, it demonstrates a high success rate in successfully repair. From the results shown in Table \ref{table:rq1_result}, only 8.0\% (4 of 50) crash bugs could not be repaired despite being able to localize the fault line. 
\vspace{-2mm}
\begin{tcolorbox}[colback=myfinding!50, colframe=white, width=\linewidth, arc=3mm, boxrule=0.5mm, left=2mm, right=2mm, top=2mm, bottom=2mm, boxsep=0mm]
\textbf{Finding 3}: The primary bottleneck in resolving code-related crash bugs lies in localization rather than repair.
\end{tcolorbox}
\vspace{-2mm}

\rev{\textbf{Code Length Impact.} To investigate the effect of code length on crash bug resolution, we compared cases with fewer than 50 lines of code to those with more than 50 lines, using the complete crash context (basic-prompt-3). For crash bugs involving complex buggy code (exceeding 50 lines), the localization and repair accuracies are 11/14 (78.6\%) and 10/14 (71.4\%), respectively. For crash bugs with simpler buggy code (fewer than 50 lines), the localization and repair accuracies are 31/36 (86.1\%) and 29/36 (80.6\%). Furthermore, only 2 out of 14 (14.3\%) cases with more than 50 lines of buggy code could be correctly localized when only the buggy code was provided. These results indicate that longer code significantly impairs both localization and repair performance.}

\vspace{-2mm}
\begin{tcolorbox}[colback=myfinding!50, colframe=white, width=\linewidth, arc=3mm, boxrule=0.5mm, left=2mm, right=2mm, top=2mm, bottom=2mm, boxsep=0mm]
\rev{\textbf{Finding 4}: Longer code leads to a poorer localization and repair ability, necessitating additional crash description information for effective resolution.}
\end{tcolorbox}
\vspace{-2mm}

\vspace{-2mm}
\subsection{Resolving Environment-related Crash Bugs} 
\label{sec:rq1_env}

\subsubsection{Design}

Mirroring the design for \codebug{} (refer to Section~\ref{sec:rq1_design}), we further explore the LLM's capability in addressing \envbug{} with varying crash description information.
Given that not all environmental crash descriptions include code snippets and that the root cause of \envbug{} may not be related to code implementation, we classify environmental crashes based on natural language descriptions and non-natural language information(\ie{} code snippets and crash information).
For each case, we first experimented with Basic-Prompt-2 and Basic-Prompt-3. If accurate repair could not be achieved with these prompts, we then
employed Basic-Prompt-4 for further investigation.

\subsubsection{Results}
\label{sec:rq1:env:result}

The LLM exhibits diminished proficiency in environment
related crash bugs resolution than code-related crash bugs, as evidenced by the lower localization and repair accuracy and more candidate solutions provided, indicating an inherent uncertainty in the LLM's bug localization.
Based on the results presented in Table \ref{table:rq1_result} and our detailed analysis of specific examples, we draw several conclusions (Finding 5-9) from different aspects.

\parabf{Overall Resolution Capability.}
Table \ref{table:rq1_result} shows that the LLM successfully identify 42 code-related crash bugs and 32 environment-related crash bugs with full crash description.
Moreover, among the accurately localized responses, 21 out of 32 (65.6\%) cannot provide a single clear solution. In some cases (e.g. crash bug 28097042~\cite{28097042} and 24305296~\cite{24305296}), the LLM gives five or more possible solutions for the same crash bug, resulting in a low accuracy. Different form \codebug{}, full stack trace is more crucial for \envbug{} diagnosis, as which contains a great deal of environment-related information.

\vspace{-3mm}
\begin{tcolorbox}[colback=myfinding!50, colframe=white, width=\linewidth, arc=3mm, boxrule=0.5mm, left=2mm, right=2mm, top=2mm, bottom=2mm, boxsep=0mm]
\textbf{Finding 5}: The LLM shows lower proficiency in resolving environment-related crash bugs compared to code-related ones due to its inability to pinpoint the root cause, resulting in multiple imprecise candidate solutions.
\end{tcolorbox}
\vspace{-3mm}

\parabf{Localization vs. Repair. }
Differing from the observations made while resolving \codebug{}, 12 instances with full crash descriptions were successfully localized but remained unresolved (only 3 instances in resolving \codebug{}). \rev{This is due to the LLM can more easily leverage error messages and crash traces to accurately localize the type of \envbug{} (such as permission, version, or IP address issues), but it often struggles to provide pecific repair steps to resolve these issues. Therefore, guiding the LLM to generate explicit and actionable repair strategies remains a key challenge for future research.}

\vspace{-3mm}
\begin{tcolorbox}[colback=myfinding!50, colframe=white, width=\linewidth, arc=3mm, boxrule=0.5mm, left=2mm, right=2mm, top=2mm, bottom=2mm, boxsep=0mm]
\textbf{Finding 6}: The LLM lacks specificity in providing solutions for environment-related crash bugs.
\end{tcolorbox}
\vspace{-2mm}

\parabf{Contextual Information Impact.}
Table \ref{table:rq1_result} shows that providing buggy code and crash information successfully identify 60\% (30 of 50) of \codebug{}. Adding crash context only results in the successful location of 12 more bugs. Addressing \envbug{} through buggy code and crash information identifies only 14\% (7 of 50) of such bugs, with only 10\% (5 of 50) successfully repaired. However, the inclusion of crash context significantly increases success rates to 64\% (32 of 50) for localization and 40\% (20 of 50) for fixing \envbug{}. This highlights the importance of crash context for \envbug{} repair, compared to \codebug{}. However, leveraging the LLM to resolve \envbug{} is challenging due to the subjective nature of crash context and its dependence on natural language, requiring more expertise from the user.

\vspace{-3mm}
\begin{tcolorbox}[colback=myfinding!50, colframe=white, width=\linewidth, arc=3mm, boxrule=0.5mm, left=2mm, right=2mm, top=2mm, bottom=2mm, boxsep=0mm]
\textbf{Finding 7}: In contrast to \codebug{}, the repair for \envbug{} relies heavily on crash context. 
\end{tcolorbox}
\vspace{-3mm}

We also observe that the LLM is unable to reason and localize the causes of a bug when the relevant environmental information is not included in the crash description. For instance, in the case of crash bug 49871007~\cite{49871007}, the exception description and stack trace only mention \textit{``Resteasy 3.1.4.Final''}. Yet, the actual library version that needed modification was \textit{``javax.ws.rs-api''}, while \textit{``Resteasy 3.1.4.Final''} is an implementation of the \textit{JAX-RS-API 2.1} specification. In such cases, the LLM fails to effectively reason and localize the cause.

\vspace{-3mm}
\begin{tcolorbox}[colback=myfinding!50, colframe=white, width=\linewidth, arc=3mm, boxrule=0.5mm, left=2mm, right=2mm, top=2mm, bottom=2mm, boxsep=0mm]
\textbf{Finding 8}: The LLM is unable to reason and localize the causes of \bug{} when the relevant environmental information is not included in the crash description.  
\end{tcolorbox}
\vspace{-2mm}

Considering the high frequency and impact of version problems in \envbug{}, which represent a significant proportion (19 out of 50 in our benchmark), we conducted a specialized analysis of these issues. We observed that the LLM typically recommends upgrading to the latest version to fix library version-related crash bugs, \eg{} bug 24305296~\cite{24305296}. However, for cases where upgrading or downgrading to a specific version of the library was the solution (\eg{} bugs 43320334~\cite{43320334} and 37771758~\cite{37771758}), the LLM couldn't provide the right version number.

\vspace{-3mm}
\begin{tcolorbox}[colback=myfinding!50, colframe=white, width=\linewidth, arc=3mm, boxrule=0.5mm, left=2mm, right=2mm, top=2mm, bottom=2mm, boxsep=0mm]
\textbf{Finding 9}: The LLM typically recommends upgrading to the latest version to solve library version-related issues.
\end{tcolorbox}
\vspace{-2mm}

\section{RQ2: Enhancing Resolution with Advanced Prompts}
\label{sec:rq2}

In this section, we explore advanced prompts to enhance the LLM's ability to address environment-related crash bugs.
To this end, we conduct experiments that investigate two key aspects of advanced prompts: prompt templates and multi-round prompts. Specifically, we first improve the design of prompt templates for single-round interactions (as detailed in Section~\ref{sec:rq2:template}), and then we enhance the resolution performance by facilitating continuous interactions in multi-round scenarios (as discussed in Section~\ref{sec:rq2_interation} and \ref{sec:active_inquiry}). This study includes a comprehensive analysis of a user study comprising 10 distinct cases.  The specific details of these cases are provided in Section~\ref{sec:rq2:cases}, while the evaluation metrics are explained in Section~\ref{sec:rq2:metric}.

\subsection{Study Cases}
\label{sec:rq2:cases}

We meticulously selected 10 diverse cases that remained unresolved despite complete crash descriptions provided in RQ1. These cases span a range of issues including dependency version conflicts, configuration settings errors, network issues, and user authority issues. 
To ensure the representativeness of the selection, cases were sampled proportionally, ensuring at least one instance from each category of issues. Detailed information for each case is presented in Table \ref{table:case}. These 10 cases will be used for specific experiments and analysis concerning each sub-problem.

\begin{table}[htp]
	\footnotesize
 \centering
	\vspace{-2mm}
	\caption{Summary of Crash Bugs Used as Case Studies }
        \vspace{-2mm}

	\label{table:case}
        \begin{adjustbox}{width=1\columnwidth}
	\begin{tabular}{|m{1.4cm}<{\centering}|m{2cm}<{\centering}|m{7cm}<{\centering}|}
		\hline
		\textbf{Bug ID} & \textbf{Crash Bug Issue} & \textbf{Solution Summary} \\ \hline
        30322026 & version issues & replace smack-java with smack-android \\ \hline
        51370703 & version issues & downgrade the version of SLF4J from 1.8.0 to 1.7.X \\ \hline
        37771758 & version issues & downgrade the version of Glassfish server from 4.1.1 to 4.1.0 \\ \hline
        60210757 & version issues & upgrade the version of to mongo-kafka-connect later than 1.0.0 \\ \hline
        49871007 & version issues & upgrade the JAX-RS-API to 2.1 \\ \hline
        43320334 & version issues & upgrade Chrome, Chrome driver and Selenium\\ \hline
        48637658 & network issues & replace the local IP address with the corresponding IP address for the Android emulator \\ \hline
        67044715 & authority issues & make sure the user has access to the table\\ \hline
        42933291 & configuration settings issues & install one of the normal arphic fonts or the texlive variant.\\ \hline
        39858254 & configuration settings issues & add the TrustStore configuration to server.xml
        \\ \hline
        
	\end{tabular}
        \end{adjustbox}
	\vspace{-5mm}
\end{table}

\subsection{Setup and Metrics}
\label{sec:rq2:metric}

We conducted the same manual examination to evaluate the correctness of the LLM's responses in terms of localization and repair as outlined in RQ1.
To mitigate the influence of the LLM's inherent randomness, we \textbf{conducted three trials for each case}. Therefore, the accuracy metric was calculated based on the proportion of successful localizations and repairs across all 30 trials.
We opted for the gpt-3.5-turbo-1106 model in alignment with RQ1.
Additionally, we rated the answers for usefulness, conciseness, and interactivity using a 4-point Likert scale~\cite{Likert1932} (1-disagree; 2-somewhat disagree; 3-somewhat agree; 4-agree) based on predefined statements (see the following). For each query, all scoring results were jointly determined by two participants through discussions.

\begin{itemize}
\item \textbf{Usefulness}: The LLM can provide useful solutions for resolving crash bugs, including a detailed explanation of the root cause of the issue and a comprehensive solution.
\item \textbf{Conciseness}: The LLM can provide clear and accurate solutions without redundant information.
\item \textbf{Interactivity}: The conversation flows well across multiple rounds, without losing sight of previous context.
\end{itemize}

\subsection{RQ2.a: Influence of Prompt Templates}
\label{sec:rq2:template}
Previous research has demonstrated that prompt templates have a significant impact on the LLM's responses~\cite{white2023prompt}. Therefore, we further explore how to improve prompt template design to enhance the LLM's performance in repairing environment-related crash bugs. This section focuses on how to design high-quality prompt templates and introduces corresponding experimental designs and results.

\subsubsection{Design}
Building on our experimentation experience and drawing inspiration from \cite{prompts}, we craft three distinct prompt templates for \envbug{}.

\begin{mdframed}[linecolor=myblue!50,linewidth=2pt,roundcorner=10pt,backgroundcolor=myyellow!20]
\small

\hspace{3.5mm}\textbf{Multi-Solution-Prompt}: Please show me all potential solutions.

\textbf{Role-Play-Prompt}: I want you to act as a fault localization and program repair expert. You will be able to provide detailed solutions to fix the given program crash. 

\textbf{Chain-of-Thought-Prompt}: Please fix it step by step.

\end{mdframed}

Firstly, we explore whether the LLM can produce the correct solution that was previously unattainable, by promoting the generation of all possible solutions through the utilization of \textbf{Multi-Solution-Prompt}.

Inspired by~\cite{dong2023self, wei2023chainofthought, zhang2022automatic}, the role-play prompt and chain-of-thought prompt may significantly enhance the LLM's efficacy in localization and repair. 
Concatenating the above prompt with \textbf{Basic-Prompt-3}, we used these prompts to conduct comparative experiments on 10 studied cases.
To mitigate any potential concerns with the randomness of the LLM, we ran three requests for each query and independently evaluated each response.

\subsubsection{Results}

\begin{table*}[htb]
	\centering
	\vspace{-3mm}
    \caption{Experimental Results of Exploring Advance Prompts}	 
    \label{table:rq2_result}
	\footnotesize
	\vspace{-2mm}
    \textbf{}	
    \begin{adjustbox}{width=1\columnwidth}
    \begin{tabular}{|m{3cm}<{\centering}|m{1.5cm}<{\centering}|m{1.5cm}<{\centering}|m{1.5cm}<{\centering}|m{1.5cm}<{\centering}|m{1.5cm}<{\centering}|m{1.5cm}<{\centering}|m{1.5cm}<{\centering}|} \hline
    
    \textbf{Prompt}  & \textbf{Rounds} & \textbf{Localization } & \textbf{Repaired} & \textbf{Solution Num.} & \textbf{Usefulness} & \textbf{Conciseness} & \textbf{Interactivity} \\  \hline
    Basic & Single & 14/30 & 3/30 & 5.8 & 1.6 & 1.5 & / \\ \hline 
    Multi-Solution & Single & 21/30 (\red{7}) & 9/30 (\red{6}) & 7.67 & 2.43 (\red{0.83})& 0.56 (\blue{0.94}) & /  \\ \hline 
    Role-Play & Single &  23/30 (\red{9}) & 10/30 (\red{7}) & 5.23 & 2.83 (\red{1.23})  & 1.87 (\red{0.37}) & /  \\ \hline 
    Chain of Thoughts & Single & 21/30 (\red{7}) & 6/30 (\red{3}) & 9.13 & 1.9 (\red{0.3}) & 0.2 (\blue{1.3}) & /  \\ \hline 
    Basic + Multi-round & Multi. & 25/30 ( \red{11}) & 16/30 ( \red{13}) & 4.77 & 2.93 (\red{1.33}) & 2.17 (\red{0.67}) & 3.37 \\ \hline 
    Role-Play + Multi-round & Multi. & 27/30 (\red{13}) & 18/30 (\red{15}) & 4.83 & 3.27 (\red{1.77}) & 2.2 (\red{0.7}) & 3.63 \\ 
    \hline 
    \end{tabular}
    \end{adjustbox}
    \vspace{-3mm}
\end{table*}

The experimental results of various prompt templates are presented in Table \ref{table:rq2_result}.
Compared to the basic prompt, the Multi-Solution-Prompt enables the LLM to generate a broader range of solutions, increasing the likelihood of finding a correct one, but resulting in reducing conciseness. Similarly, the Chain-of-Thought-Prompt introduces substantial redundancy, leading to a low conciseness score of 0.2, without significantly improving the LLM’s ability to derive a more accurate end-to-end solution.
Compared to the Multi-Solution and Chain-of-Thought Prompts, which enhance localization and repair effectiveness but sacrifice conciseness, the Role-Play-Prompt excels in single-round interactions. It outperforms the basic prompt with 0.23 higher accuracy in repairs, a 1.23 higher usefulness score, and improved conciseness.

Furthermore, the Role-Play-Prompt improves the LLM’s ability to identify the root causes of crash bugs. For instance, while the first two prompts recognized issues related to the Smack version in crash bug 30322026, they failed to identify the mismatch between the Smack version and the Android project as the underlying cause, which the responses of Role-Play-Prompt successfully pinpointed. Thus, we conclude:

\vspace{-2mm}
\begin{tcolorbox}[colback=myfinding!50, colframe=white, width=\linewidth, arc=3mm, boxrule=0.5mm, left=2mm, right=2mm, top=2mm, bottom=2mm, boxsep=0mm]
\textbf{Finding 10}: Role-play prompts significantly improve the LLM's ability to localize and repair crash bugs.
\end{tcolorbox}
\vspace{-4mm}

\subsection{RQ2.b: Impact of Multi-round Prompts}
\label{sec:rq2_interation}
We investigate the impact of multi-round interactions on improving the LLM’s ability to resolve crash bugs, hypothesizing that iterative interactions can refine its responses and enhance overall performance.

\subsubsection{Design}

To evaluate this hypothesis, we conduct experiments assessing the effectiveness of providing feedback on the LLM’s responses across two different scenarios.

\begin{itemize}[itemsep=2pt,topsep=0pt,parsep=0pt]
    \item The initial answer did not accurately localize the root cause of the \bug{} (\eg{} crash bug 37771758).
    
    \item The initial answer identified the root cause of the \bug{}, but the solution provided was too general and did not offer specific resolution steps (\eg{} crash bugs 51370703 and 30322026).
\end{itemize}

For the first scenario, we devise the New-Solution-Prompt to request new solutions; for the second scenario, we use the Refinement-Prompt for the specific details of the solution.
During our attempts, we found that Refinement-Prompt can obtain solution details for most issues. However, for issues related to library versions, Refinement-Prompt was insufficient in obtaining specific version numbers that needed to be upgraded or downgraded. Therefore, we designed specific prompts for such problems that require targeted questioning. For each crash bug, we employed both Basic-Prompt and Role-Play-Prompt in our experiments.

\begin{mdframed}[linecolor=myblue!50,linewidth=2pt,roundcorner=10pt,backgroundcolor=myyellow!20]
\small

\hspace{3.5mm}\textbf{New-Solution-Prompt:} I have tried the solution above, but the issue remains. Can you give me some new solutions?

\textbf{Refinement-Prompt:} 
Please provide more detailed information regarding the solutions mentioned above. 

\textbf{Version-Prompt:} Which version of [Library 1] is compatible with my project/ [Library 2] version?
\end{mdframed}

\subsubsection{Results}

Based on the specific analysis of the study cases and the results shown in Table~\ref{table:rq2_result}, we can draw the following conclusions (Findings 11-13), each with a detailed analysis.

Table \ref{table:rq2_result} illustrates that continuous interaction enhances the model's performance in both localization and repair accuracy, as well as in scoring for usefulness and conciseness, compared to single-round interactions. Notably, even with the basic prompt in continuous interaction, the results surpass those of the most effective single-round role-play prompt. This underscores the effectiveness of continuous interaction.

\vspace{-3mm}
\begin{tcolorbox}[colback=myfinding!50, colframe=white, width=\linewidth, arc=3mm, boxrule=0.5mm, left=2mm, right=2mm, top=2mm, bottom=2mm, boxsep=0mm]
\textbf{Finding 11}: Continuously interacting with the LLM can help to further improve its ability to localize and repair \bug{}.
\end{tcolorbox}
\vspace{-2mm}

The experiments in Table \ref{table:rq2_result} show that utilizing the Role-Play-Prompt for multi-turn interactions outperforms the basic prompt across all metrics. This suggests that the Role-Play-Prompt helps the LLM better understand the context and generate more accurate responses. Additionally, it significantly enhances the interaction score from 3.37 to 3.63. This improvement can be attributed to the Role-Play-Prompt aiding the LLM in maintaining contextual consistency and avoiding the repetition of previous solutions.

\vspace{-2mm}

\begin{tcolorbox}[colback=myfinding!50, colframe=white, width=\linewidth, arc=3mm, boxrule=0.5mm, left=2mm, right=2mm, top=2mm, bottom=2mm, boxsep=0mm]
\textbf{Finding 12}: The use of Role-Play-Prompt significantly improves the LLM's ability to maintain contextual coherence and reduces instances of forgetting previous context, resulting in better resolving and interaction performance.
\end{tcolorbox}
\vspace{-2mm}

Based on our trials with crash bug 51370703, we provide the following conclusion as a supplement to finding 8.

\vspace{-2mm}
\begin{tcolorbox}[colback=myfinding!50, colframe=white, width=\linewidth, arc=3mm, boxrule=0.5mm, left=2mm, right=2mm, top=2mm, bottom=2mm, boxsep=0mm]
\textbf{Finding 13}: the LLM appears to have knowledge about library versions that are discussed in official documentation. However, it requires targeted questioning to activate the corresponding knowledge.
\end{tcolorbox}
\vspace{-2mm}

For instance, during continuous interaction, a vague inquiry such as \textit{``Which version of SLF4J should I use?''} lead the LLM to merely suggest upgrading to the latest available version or recommend version 1.8, as referenced in the bug report. In contrast, posing a more targeted question such as \textit{``Which version of SLF4J is compatible with Logback 1.2.3 in my project?''} can prompt the LLM to retrieve specific knowledge that \textit{Logback version 1.2.3 explicitly requires the slf4j-api version 1.7.x}.

\subsection{\rev{RQ2.c: Enhancing Multi-round Prompts with Self-planning}}
\label{sec:active_inquiry}
From the analysis above, we observe that even with advanced prompts such as the Role-Play Prompt and Chain-of-Thought prompt, the LLM struggles to accurately identify the root cause of a crash bug in a single attempt, limiting its ability to provide targeted solutions.
As discussed in Section~\ref{sec:rq2_interation} and Finding 13, while the LLM possesses the necessary knowledge to address many crash bugs, effectively utilizing this knowledge requires precise strategies. Continuous interaction and targeted questioning rely on the questioner’s ability to discern the true cause from multiple potential solutions, imposing high demands on the user.
Furthermore, the lack of sufficient contextual information from the questioner constrains the LLM's ability to generate targeted solutions (Finding 8). Additionally, questioners may struggle to identify the relevant aspects of the broader environment-related context necessary for accurately diagnosing the crash bug.

To address these challenges, we empower the LLM to actively investigate potential crash-causing environmental factors through continuous proactive inquiry, guiding the resolution process step by step. \rev{By leveraging the LLM's self-planning capability~\cite{jiang2023self}, we propose an \textbf{active inquiry prompt} that guides the LLM in constructing an investigative plan. This plan examines various environmental factors that may contribute to crashes, prioritizes them based on diagnostic relevance, and acquires necessary information by posing targeted questions to the user. Although iterative feedback-based prompting has been explored in prior software engineering research~\cite{jin2023inferfix, bouzenia2024repairagent}, the key innovation of the active inquiry prompt lies in its ability to stimulate the LLM to ask proactive and targeted questions. Through this iterative questioning process, the LLM progressively refines its understanding, gradually narrows down the root cause, and ultimately provides precise solutions. Moreover, this approach proves particularly beneficial for novice users who may lack specialized expertise, enabling them to more effectively resolve environment-related crash bugs.}

\subsubsection{Design and Results}
We used the role-play prompt to enable the LLM for active inquiry. We refined the questioning method to avoid overly general queries and prevent overwhelming developers with multiple questions simultaneously. This included specifying certain question types and providing an example to guide the LLM's questioning toward more targeted and actionable suggestions. The resulting prompt stimulates the LLM's proactive questions, facilitating self-planning.

\begin{mdframed}[linecolor=myblue!50,linewidth=2pt,roundcorner=10pt,backgroundcolor=myyellow!20]
\small
\hspace{3.5mm}\textbf{ActiveQ-Prompt:} 
[Role-Play-Prompt] Additionally, please use the Socratic method of questioning to aid in accurate diagnosis. Note that the information in the question should be as specific as possible. For example, ...[AskOneQ-Prompt]

\textbf{AskOneQ-Prompt}: Note that only ask one question at a time, starting with the question you think is the most important.

\end{mdframed}

By using this method, the LLM is able to ask one question at a
time based on its priority, leading developers to targeted solutions.
Utilizing the crash bug numbered 30322026 as an illustrative case, the LLM posed the following inquiry with the utilization of ActiveQ-Prompt and AskOneQ-Prompt:
\begin{mdframed}[linecolor=myblue!50,linewidth=2pt,roundcorner=10pt,backgroundcolor=myanswer!20]
\small
\textit{My first question is: What is the version of the Smack library that you are using in this project?}
\end{mdframed}

After providing the corresponding information, partial the LLM's response would be as follows:
\begin{mdframed}[linecolor=myblue!50,linewidth=2pt,roundcorner=10pt,backgroundcolor=myanswer!20]
\small
\textit{To fix this issue, you can try using a Smack library version that is specifically designed for Android,such assmack-android-extensions:4.4.4`orsmack-android:4.4.4`.}
\end{mdframed}

Based on the response, it is evident that the LLM accurately pinpointed the underlying cause of the bug, which pertains to the incompatibility of the employed smack-java7:4.1.0 library with the Android environment. Furthermore, the LLM suggested a correct library version as the detailed solution. 
\vspace{-2mm}
\begin{tcolorbox}[colback=myfinding!50, colframe=white, width=\linewidth, arc=3mm, boxrule=0.5mm, left=2mm, right=2mm, top=2mm, bottom=2mm, boxsep=0mm]

\textbf{Finding 14}: Employing active inquiry prompts to stimulate proactive questions for self-planning enhances the LLM's ability to proactively guide the repair process. This aids in identifying the root causes of crash bugs, proving particularly beneficial for novice users.

\end{tcolorbox}
\vspace{-4mm}

\section{Methodology: \app{}}

Due to the inherent non-determinism of \envbug{}, their symptoms manifest through complex interactions between software components and runtime configurations, presenting significant challenges in developing end-to-end solutions. Based on our experimental findings, we propose \app{}, a novel approach for addressing environment-related crash bugs through continuous interaction with LLMs. 

\begin{figure}[htb]
	\centering
        \vspace{-3mm}
	\includegraphics[width=0.60\columnwidth]{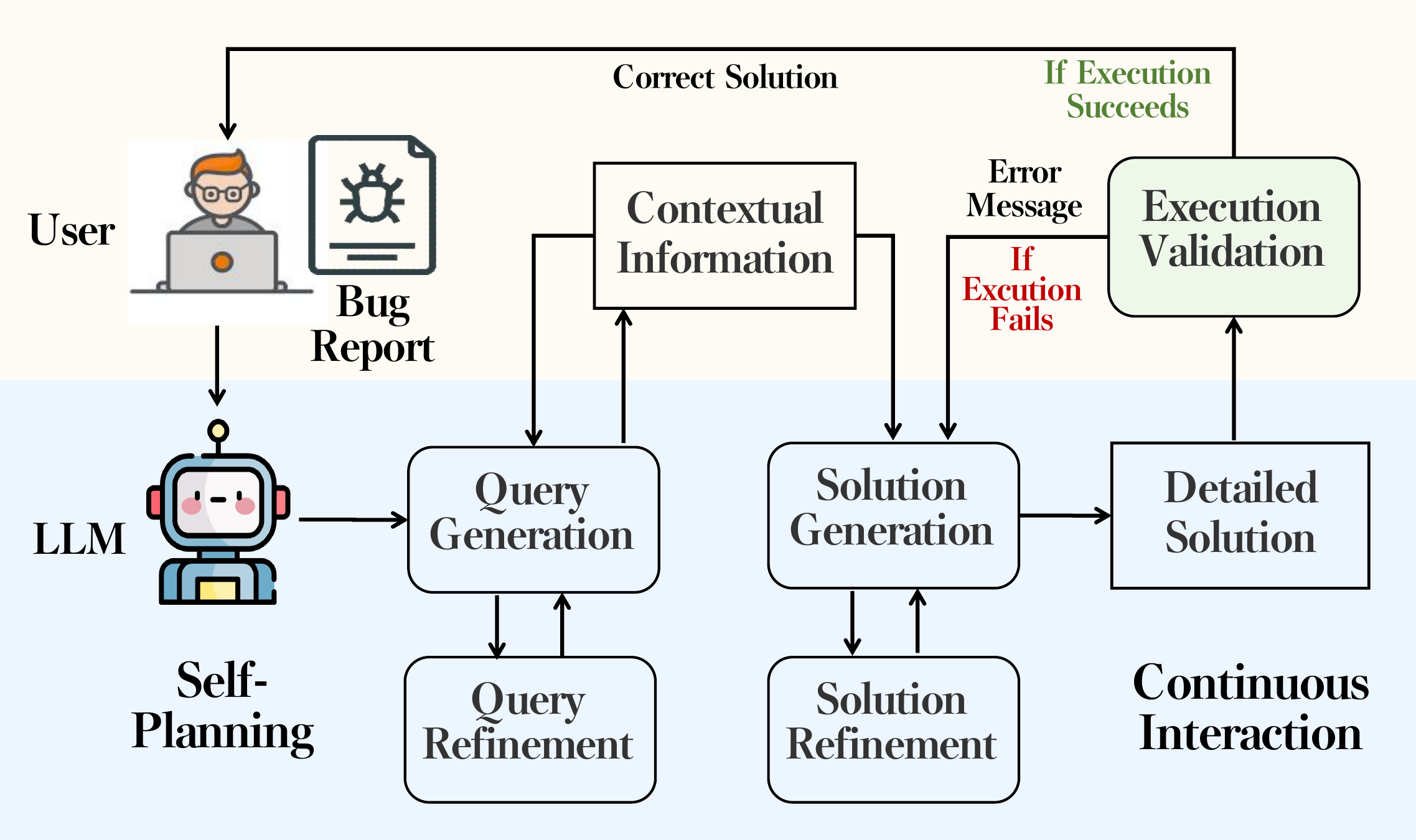}
        \vspace{-2mm}

	\caption{Overview of \app{}}
	\label{fig:overview}
    \vspace{-3mm}
\end{figure}

Figure \ref{fig:overview} illustrate the overview of our approach. The proposed framework operates through a continuous interaction cycle between the user and the LLM, comprising three main phases: \textbf{Phase 1: Contextual Information Completion via LLM Self-Planning.}
Given the bug report, this phase utilizes the LLM's self-planning capabilities to automatically generate and iteratively refine queries. These queries aim to gather essential contextual information that enriches the original bug report.
\textbf{Phase 2: Solution Generation.} Utilizing the enriched context from the previous phase and the initial bug report, the LLM generates potential solutions. This generation process involves iterative refinement prompts designed to progressively yield a specific and actionable detailed solution.
\textbf{Phase 3: Solution Validation.} In the final phase, we evaluate the correctness of the detailed solution  through execution validation. If the execution succeeds, the verified correct solution is returned directly to the user. Conversely, if the execution fails, the LLM iteratively refines and generates a revised solution.

\begin{figure}[htb]
	\centering
        \vspace{-3mm}
	\includegraphics[width=0.82\columnwidth]{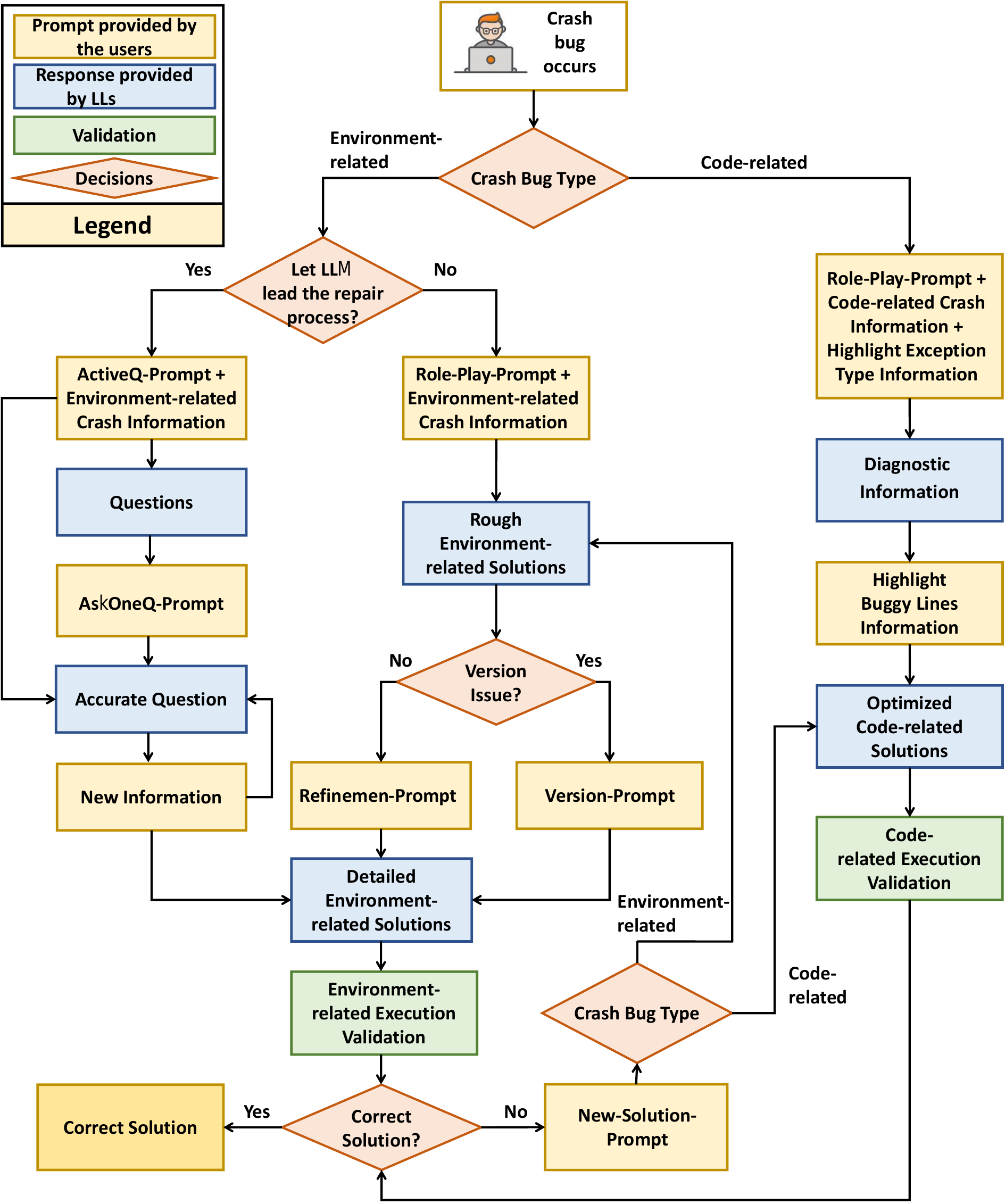}
        \vspace{-2mm}

	\caption{Detailed Workflow of \app{} with Specific Prompts}
	\label{fig:overview_prompt}
    \vspace{-3mm}
\end{figure}

Figure~\ref{fig:overview_prompt} illustrates the detailed workflow with specific prompts utilized by \app{}. To enhance the generalizability of \app{}, we extend its application to effectively handle both environment-related and code-related crash bugs.

\parabf{Concept Explanations.} \textbf{Prompt provided by the users} is in a yellow box with crash details and instructions. The details of different prompts can be referred to the prompt templates listed in RQ1 and RQ2. \textbf{Response provided by LLMs} is in a blue box and can be categorized into questions and solutions. Questions may be multiple, but highlighting the question requirement will help the LLM generate a specific question (seeing section \ref{sec:active_inquiry}). Solutions can be rough or detailed (including optimized solution) with specific steps to fix the crash bugs. \textbf{Validation} is the repair attempt result in green box. \textbf{Decision} in diamond-shaped box indicating a selection, which based on manual judgment currently.

\parabf{Code-related Crash Bug Resolution.}The main challenge for resolving \codebug{} lies in accurately diagnosis the root cause of crash bugs (refer to finding 3). Therefore, utilizing a Chain-of-Thought strategy to divide the resolution into diagnosis and repair stages is beneficial. This strategy entails deploying prompts to guide LLMs in identifying the buggy line, subsequently emphasizing the pinpointed information for efficient repair. Additionally, accentuating the exception type and providing a comprehensive crash context are instrumental in amplifying the effectiveness of the resolving (refer to finding 2).

\parabf{Environment-related Crash Bug Resolution.}
For \envbug{} resolving, users can choose to let LLMs guide the repair process. This is helpful for beginners with limited crash bug knowledge, while seasoned developers may not need LLMs' active guidance. 
Moreover, version issues present unique challenges as discussed in section \ref{sec:rq1_env} and finding 13. If users choose not to rely on LLMs' guidance, they should carefully examine the initial solution to identify any version issues. Version prompts should be utilized to elicit specific details in such cases. 

\parabf{Validation Process}. 
Validation involves receiving LLMs' solutions and confirming whether they fix the crash bug. If not, new solution prompts are used, with a maximum retry limit (\eg 3 attempts in our setting).

\section{Evaluation of \app{}}
To assess the effectiveness and generalizability of \app{}, we conducted comprehensive assessments for resolving crash bugs utilizing diverse LLMs.
Initially, we conducted experiments by integrating \app{} based on GPT-3.5, which we employed in RQ1 and RQ2 (as detailed in Section~\ref{sec:rq3:a}). Since \app{} was developed based on the insights from RQ1 and RQ2, our objective was to determine its efficacy in addressing the identified challenges and previously unresolved crash bugs from those studies.
Moreover, we further evaluated the generalizability of \app{} by applying it to a range of state-of-the-art LLMs, encompassing both open-source and closed-source models (Section~\ref{sec:rq3:b}), \rev{as well as by testing its effectiveness on crash bugs from different programming languages (Section~\ref{sec:rq3:c}).}

\subsection{RQ3: Effectiveness of \app{}}
\label{sec:rq3:a}
We assess the effectiveness of \app{} in resolving crash bugs using GPT-3.5 and compare it to a baseline scenario that employed with a basic prompt.

\subsubsection{Participants.} We invited two participants, each with over two years of Java programming experience, to conduct experiments and evaluations independently. Their evaluations were based on the criteria for the correctness of localization and repair as RQ1-defined metrics. In cases of disagreement, an author involved in RQ1 and RQ2 provided arbitration.

\subsubsection{Benchmark.}
From the initial 100-crash-bug benchmark in Section \ref{sec:benchmark}, we deliberately chose unrepaired instances (see Table \ref{table:rq1_result}) despite receiving complete crash descriptions. This selection forms a refined benchmark, comprising 11 \codebug{} and 30 \envbug{} crashes, totaling 41 cases.

\subsubsection{Metrics.}
The experiment's metrics align with those presented in Section \ref{subsec:rq1_result}. Additionally, to assess the effectiveness of continuous interaction, we record the number of interaction rounds required to achieve the final correct localization or repair result. \rev{To mitigate any potential concerns with the randomness of the LLM, each query was executed three times using both \app{} and the baseline methods. For the calculation of localization and repair accuracy, a case is considered successfully resolved only if at least two out of the three runs yield a correct result.}

\begin{table*}[htb]
    \centering
    \vspace{-5mm}
    \caption{Effectiveness of \app{} across different LLMs}	 
    \label{table:final_result}
     \vspace{-3mm}
    \footnotesize
    \begin{adjustbox}{width=\textwidth}
    \begin{tabular}{|m{1.8cm}<{\centering}|m{2cm}<{\centering}|m{2cm}<{\centering}|m{1.4cm}<{\centering}|m{1.4cm}<{\centering}|m{1.4cm}<{\centering}|m{2cm}<{\centering}|m{2cm}<{\centering}|} 
    \hline
    \textbf{Model} & \textbf{Type}& \textbf{Prompt} &  \textbf{Localization} &  \textbf{Repair} & \textbf{Multi. Rounds} & \textbf{Localization Accuracy} & \textbf{Repair Accuracy}\\  \hline
    \multirow{4}{*}{GPT-3.5} & \multirow{2}{*}{Code Related} & Basic Prompt & 3/11 & 0/11 & / & 27.3\% & 0.0\% \\ \cline{3-8}
    & & \app{} & 7/11  (\red{4}) & 4/11  (\red{4}) & 3 & 63.6\% (\red{36.3\%}) & 36.4\% (\red{36.4\%})\\ \cline{2-8}
    & \multirow{2}{*}{Env. Related}  & Basic Prompt & 12/30 & 0/30 & / & 40.0\% &  0.0\%\\ \cline{3-8}
    & & \app{} & 25/30 (\red{13}) & 16/30 (\red{16}) & 14 & 83.3\% (\red{43.3\%}) &  53.3\% (\red{53.3\%})\\ \hline
    \multirow{4}{*}{GPT-4} & \multirow{2}{*}{Code Related} & Basic Prompt & 8/11 & 5/11 &  / & 72.7\% & 45.5\% \\ \cline{3-8}
    & & \app{} & 9/11  (\red{1}) & 6/11 (\red{1}) & 1 & 81.8\% (\red{9.1\%}) & 54.5\% (\red{9.1\%}) \\ \cline{2-8}
    & \multirow{2}{*}{Env. Related}  & Basic Prompt & 21/30 & 10/30 & / & 70.0\% &33.3\%  \\ \cline{3-8}
    & & \app{} & 25/30 (\red{4}) & 21/30 (\red{11}) & 9 &83.3\% (\red{13.3\%})  &70.0\% (\red{36.7\%})   \\ \hline
    \multirow{4}{*}{Claude} & \multirow{2}{*}{Code Related} & Basic Prompt & 5/11 & 4/11 & / & 45.5\% & 36.4\% \\ \cline{3-8}
    & & \app{} & 6/11  (\red{1}) & 5/11  (\red{1}) & 1 & 54.5\% (\red{9.1\%}) & 45.5\% (\red{9.1\%}) \\ \cline{2-8}
    & \multirow{2}{*}{Env. Related}  & Basic Prompt & 12/30 & 5/30 & / & 40.0\% & 16.7\%\\ \cline{3-8}
    & & \app{} & 18/30 (\red{6}) & 13/30 (\red{8}) & 8 & 60.0\% (\red{20.0\%})  & 43.3\% (\red{26.6\%})  \\ \hline
    \multirow{4}{*}{Codellama} & \multirow{2}{*}{Code Related} & Basic Prompt & 4/11 & 2/11 & / & 36.4\% & 18.2\%   \\ \cline{3-8}
    & & \app{} & 6/11 (\red{2}) & 4/11  (\red{2}) & 1 & 54.5\% (\red{18.1\%}) & 36.4\% (\red{18.2\%}) \\ \cline{2-8}
    & \multirow{2}{*}{Env. Related}  & Basic Prompt & 7/30 & 2/30 & / & 23.3\% &6.7\% \\ \cline{3-8}
    & & \app{} & 12/30 (\red{5}) & 9/30 (\red{7}) & 7 & 40.0\% (\red{16.7\%}) & 30.0\% (\red{23.3\%}) \\ \hline
    \multirow{4}{*}{Deepseek-r1} & \multirow{2}{*}{Code Related} & Basic Prompt & 8/11 & 7/11 & / & 72.7\% & 63.6\% \\ \cline{3-8}
    & & \app{} & 11/11  (\red{3}) & 11/11 (\red{4}) & 2 & 100.0\% (\red{27.3\%}) & 100.0\% (\red{36.4\%}) \\ \cline{2-8}
    & \multirow{2}{*}{Env. Related}  & Basic Prompt & 19/30 & 13/30 & / & 63.3\% & 43.3\% \\ \cline{3-8}
    & & \app{} & 28/30 (\red{9}) & 26/30 (\red{13}) & 12 & 93.3\%(\red{30.0\%})  & 86.7\%(\red{43.4\%})   \\ \hline
    
     \multirow{4}{*}{Qwen3-coder} & \multirow{2}{*}{Code Related} & Basic Prompt & 8/11 & 7/11 & / & 72.7\% & 63.6\% \\ \cline{3-8}
    & & \app{} & 10/11  (\red{2}) & 9/11 (\red{2}) & 1 & 90.9\% (\red{18.2\%}) & 81.8\% (\red{18.2\%}) \\ \cline{2-8}
    & \multirow{2}{*}{Env. Related}  & Basic Prompt & 15/30 & 11/30 & / & 50.0\% & 36.7\% \\ \cline{3-8}
    & & \app{} & 26/30 (\red{11}) & 24/30 (\red{13}) & 10 & 86.7\%(\red{36.7\%})  & 80.0\%(\red{43.3\%})   \\ \hline
    \end{tabular}
    \end{adjustbox}
    \vspace{-4mm}
\end{table*}

\subsubsection{Results.} The results demonstrate significant improvements in resolving both \codebug{} and \envbug{}. \rev{Specifically, localization accuracy increased by 36.3\% and 43.3\%, while repair accuracy improved by 36.4\% and 53.3\%, respectively.}

In \codebug{} resolution, 4 additional crash bugs were effectively localized and repaired. Three among these were repaired by emphasizing the buggy line in subsequent interactions, while highlighting the exception type in initial crash description facilitated one's localization.
Concerning \envbug{}, the improvement was more significant: 13 more bugs were accurately localized, and 16 more were successfully repaired. Furthermore, 14 crash bugs were resolved through iterative interactions, highlighting the efficacy of our \app{} framework.

\subsection{RQ4.a: Generalizability of \app{} across LLMs}
\label{sec:rq3:b}
We evaluate \app{}'s generalizability on \rev{five} additional LLMs, employing the same benchmarks and metrics as in RQ3 (Section~\ref{sec:rq3:a}).

\subsubsection{Studied LLMs}
\label{studied_llms}
\rev{We select five state-of-the-art LLMs: two closed-source models (GPT-4~\cite{openai2023gpt4} and Claude~\cite{claude}) and three open-source models (Codellama-34b~\cite{codellama}, DeepSeek-R1~\cite{deepseek} and Qwen3-Coder-Instruction~\cite{qwen}). }
To ensure experimental consistency, we maintain a uniform approach by conducting manual interactions through a web interface for all the studied LLMs.

\subsubsection{Results}
\rev{\textbf{Overall Performance.} Table \ref{table:final_result} demonstrates \app{}'s remarkable generalizability, effectively improving resolution for both \codebug{} and \envbug{} across various closed-source and open-source LLMs when compared to the basic prompt, with substantial enhancements ranging from 9.1\% to 43.3\% in localization and 9.1\% to 53.3\% in repair.}
\rev{\parabf{Localization vs. Repair.} Overall, the improvement in the repair stage is more pronounced than in the localization stage, particularly for environment-related crash bugs. This highlights the effectiveness of \app{} in addressing the challenge where LLMs often struggle to provide specific  repair steps for environment-related crash bugs (see Finding 6).}
\rev{\parabf{Weaker LLMs vs. Stronger LLMs.} All models benefit significantly from the introduction of \app{}. For code-related crash bugs, some stronger models (such as DeepSeek-R1) already achieve high performance with the basic prompt, sometimes even outperforming weaker models (such as GPT-3.5 and CodeLlama) using \app{}. However, for environment-related crash bugs, weaker models using \app{} still surpass the performance of stronger models using only the basic prompt. This indicates that resolving environment-related crash bugs remains challenging even for state-of-the-art LLMs, underscoring the value of \app{}. Additionally, except for CodeLlama, all models show greater improvement in resolving \envbug{} than \codebug{}, which is consistent with our focus on environment-related bugs in Section~\ref{sec:rq2} and leads to more substantial enhancements in this area.}

\subsection{\rev{RQ4.b: Generalizability of \app{} across Programming Languages}}
\label{sec:rq3:c}
\rev{We constructed an expanded small-scale multilingual dataset to evaluate the generalizability of IntDiagSolver across different programming languages, including Java, C/C++, and Python, using the same metrics as in RQ3 (Section~\ref{sec:rq3:a}).}

\subsubsection{\rev{Benchmark.}}

\rev{To build this multilingual dataset, we invited the same two participants from Section~\ref{sec:rq3:b} to follow the data collection procedures described in Section~\ref{sec:benchmark}. We collected a total of 42 environment-related crash bugs spanning Java, C/C++, and Python for further generalization evaluation (with 10 cases for Java, 19 for Python, and 6 for C/C++). To further assess the effectiveness of LLMs on previously unseen data, all 42 crash bugs were sourced from Stack Overflow posts published after 2024.}

\subsubsection{\rev{Studied LLMs.}}

\rev{For this generalization experiment, we selected two representative advanced models: 
a closed-source model (i.e., GPT-4o~\cite{gpt4o})
and a open-source model(i.e., DeepSeek-R1~\cite{deepseek}). }

\subsubsection{\rev{Results.}} 
\rev{\parabf{Overall Performance.} As shown in Table~\ref{table:rq4}, \app{} consistently demonstrates significant improvements over the baselines across all programming languages. Specifically, localization accuracy improved by 12.0\% to 16.6\%, and repair accuracy improved by 19.0\% to 21.4\% across the two advanced LLMs.
\parabf{Impact of Programming Language.} Across all three programming languages, \app{} shows varying degrees of improvement when combined with both LLMs. Notably, both models achieved the most significant gains in resolving Python crash bugs, particularly in repair accuracy.
\parabf{Performance on Unseen Data.} Overall, the comparative results between \app{} and the baseline on this unseen dataset closely mirror those observed in RQ4.a. The application of \app{} results in a significant performance improvement, which not only confirms its effectiveness on previously unseen data, but also demonstrates that the experimental results in RQ1–RQ4 are not influenced by data contamination.}

\begin{table*}[htb]
    \centering
    \vspace{-5mm}
    \caption{Effectiveness of \app{} across different programming languages}	 
    \label{table:rq4}
    \vspace{-3mm}
    \footnotesize
    \begin{adjustbox}{width=\textwidth}
    \begin{tabular}{|m{1.8cm}<{\centering}|m{2cm}<{\centering}|m{1.8cm}<{\centering}|m{1.6cm}<{\centering}|m{1.6cm}<{\centering}|m{2.2cm}<{\centering}|m{2.2cm}<{\centering}|} 
    \hline
    \textbf{Model} & \textbf{Prompt} & \textbf{Language} & \textbf{Localization} & \textbf{Repair} & \textbf{Localization Accuracy} & \textbf{Repair Accuracy}\\  
    \hline

    \multirow{8}{*}{GPT-4o} 
     & \multirow{4}{*}{Basic Prompt} & Python & 15/20 & 11/20 & 75.0\% & 55.0\% \\ 
     &  & Java & 8/13 & 7/13 & 61.5\% & 53.8\% \\ 
     &  & C/C++ & 5/9 & 4/9 & 55.6\% & 44.4\% \\ 
     &  & All & 28/42 & 22/42 & 66.7\% & 52.4\% \\ 
     \cline{2-7}
     & \multirow{4}{*}{\app{}} & Python & 19/20  (\red{4}) & 18/20  (\red{7}) & 95.0\% (\red{20.0\%}) & 90.0\% (\red{35.0\%})\\ 
     &  & Java & 10/13  (\red{2}) & 7/13  (\textcolor{red}{+0}) & 76.9\% (\red{15.4\%}) & 53.8\% (\textcolor{red}{+0\%})\\ 
     &  & C/C++ & 6/9  (\red{1}) & 5/9  (\red{1}) & 66.7\% (\red{11.1\%}) & 55.6\% (\red{11.2\%})\\ 
     &  & All & 35/42  (\red{7}) & 30/42  (\red{8}) & 83.3\% (\red{16.6\%}) & 71.4\% (\red{19.0\%})\\  
    \hline

    \multirow{8}{*}{DeepSeek-r1} 
     & \multirow{4}{*}{Basic Prompt} & Python & 16/20 & 11/20 & 80.0\% & 55.0\% \\ 
     &  & Java & 8/13 & 6/13 & 61.5\% & 46.2\% \\ 
     &  & C/C++ & 5/9 & 5/9 & 55.6\% & 55.6\% \\ 
     &  & All & 29/42 & 22/42 & 69.0\% & 52.4\% \\ 
     \cline{2-7}
     & \multirow{4}{*}{\app{}} & Python & 18/20  (\red{2}) & 18/20  (\red{7}) & 90.0\% (\red{10.0\%}) & 90.0\% (\red{35.0\%})\\ 
     &  & Java & 10/13  (\red{2}) & 8/13  (\red{2}) & 76.9\% (\red{15.4\%}) & 61.5\% (\red{15.3\%})\\ 
     &  & C/C++ & 6/9  (\red{1}) & 5/9  (\red{0}) & 66.7\% (\red{11.1\%}) & 55.6\% (\red{0\%})\\ 
     &  & All & 34/42  (\red{5}) & 31/42  (\red{9}) & 81.0\% (\red{12.0\%}) & 73.8\% (\red{21.4\%})\\  
    \hline

    \end{tabular}
    \end{adjustbox}
    \vspace{-4mm}
\end{table*}

\section{Related Work}
\subsection{LLMs for Software Engineering}

In recent years, LLMs have gained considerable attention across various research domains, including mathematics~\cite{davis2023mathematics, frieder2023mathematical}, education \cite{gilson2022education, baidoo2023education}, and natural language processing \cite{bang2023nlp, jiao2023nlp}.
In the field of software engineering (SE), LLMs have demonstrated their potential by being applied to a wide range of tasks, from code generation \cite{poesia2022codegeneration, zeng2022codegeneration, du2023classeval} and code summarization \cite{wang2020codesummarization, 23codesummarization} to software maintenance tasks, including vulnerability detection \cite{sun2023gpt, zhang2023promptenhanced}, test generation \cite{yuan2023manual, yu2023llm}, and program repair \cite{msr21apr, arXiv23AutomaticBugFixing, arXiv23ConversationalProgramRepair, jin2023inferfix, icse2023apr}. This broad SE application stems from their robust training on extensive code and text data, enhancing both linguistic and code comprehension.

\subsection{LLMs for Resolving Crash Bugs}
Software crashes pose an enduring challenge in software development, driving research in areas such as crash reproduction~\cite{DBLP:conf/icse/JinO12, DBLP:journals/smr/NayrollesHTL17}, crash localization~\cite{DBLP:conf/issta/WuZCK14, DBLP:conf/msr/WangKZ13, DBLP:conf/sigsoft/0001R18}, and crash repair~\cite{ase15crashresolve, icse21crasolver, fse20masetro}. Efforts to tackle code-related crash bugs have resulted in research on automatic localization~\cite{cao2020fl, ase2021fl, icse22fl} and repair~\cite{msr21apr, arXiv23chatgptrepair, arXiv23ConversationalProgramRepair}. In contrast, environment-related crash bugs, due to their diverse origins, often rely on solutions from online Q\&A forums like SO~\cite{icse21crasolver, fse20masetro, icst22masetroextend, du2023kg4crasolver}.

Recent surveys on LLMs in software engineering~\cite{hou2023large, yuan2023evaluating} have explored their applications, performance, and challenges, including bug localization and program repair. Preliminary studies~\cite{arXiv23AutomaticBugFixing, arXiv23ConversationalProgramRepair, jin2023inferfix} on ChatGPT's potential in addressing code-related bugs have primarily focused on simple scenarios with fixed prompts. In contrast, our study: (i) diverges from existing work by categorizing the resolution process into two stages: localization and repair, (ii) comprehensively assesses ChatGPT's effectiveness in resolving real-world crash bugs sourced from SO, (iii) explores the effects of different prompts in various interaction strategies, and (iv) notably extends beyond code-related problems to include crash bugs caused by external environmental factors, an aspect that has received limited attention in prior research.

\vspace{-2mm}

\subsection{Prompt Engineering for Software Engineering}
Prompt engineering is an emerging discipline that optimizes prompts for various applications of LLMs across different domains. Methodologies in this field include Few-Shot Prompting~\cite{brown2020language}, Chain-of-Thought Prompting (CoT)~\cite{wei2023chainofthought, zhang2022automatic}, Tree-of-Thoughts (ToT)~\cite{yao2023tree}, and Knowledge Prompting~\cite{liu2022generated}.
In SE domain, prompt engineering enhances LLMs in various tasks. Researchers have used CoT Prompting to improve code generation~\cite{liu2023improving, li2023structured}, explored argumentation and task decomposition strategies for better unit test generation~\cite{yuan2023manual}, and combined CoT Prompting with static analysis for more efficient vulnerability detection~\cite{zhang2023promptenhanced}. 
In this study, we explored various prompt templates and advanced techniques to effectively resolve crash bugs. We further propose \app{}, employing Knowledge Prompting to resolve crash bugs. It breaks the process into localization and repair with multi-turn interactions involving LLMs, following the principles of CoT Prompting. 

\section{Threats of Validity}

The empirical study used a dataset of 100 Java-related crash bugs from SO, which might limit generalizability. \rev{To address this, we constructed an expanded small-scale multilingual dataset to evaluate the generalizability of our method across different programming languages in RQ4. Another threat arises from the fact that our evaluation compares LLM-generated answers against the accepted answers on Stack Overflow, without additional empirical validation. Nevertheless, the accepted answers were originally verified for correctness by the question askers themselves, providing a reasonable basis for reliability. Moreover, due to the lack of sufficient contextual information in most Stack Overflow threads, reproducing the crashes for empirical validation was infeasible, so we followed the same evaluation methodology adopted in prior studies~\cite{accanswer1, accanswer2, accanswer3, accanswer4}. For future work, we plan to manually collect and construct a dedicated benchmark that includes sufficient contextual information to enable full empirical validation of crash reproduction and resolution.}

In our exploratory study (RQ1 and RQ2), employing a specific LLM (i.e., GPT-3.5) may limit the generalizability of our findings and the broader applicability of the proposed \app{} approach. To mitigate this concern, in RQ3, we evaluated our approach using various state-of-the-art LLMs, reinforcing our initial findings and indicating wider applicability across different models.
Another potential limitation of RQ2 stems from exploring interactions using only 10 cases, which may introduce selection bias. However, subsequent practical evaluations demonstrated promising effectiveness across a broader range of cases, as shown in Table~\ref{table:final_result}. Due to the inherently stochastic nature of LLM outputs, there is an inherent risk to the reliability and validity of experimental results. To address this, we conducted multiple trials and rigorously documented all interactions with LLMs to ensure transparency and reproducibility~\cite{replication_package}.

An additional threat involves potential data leakage from the Stack Overflow benchmark dataset. Nevertheless, our investigations in RQ1 and RQ2 revealed that model performance predominantly depends on contextual variations of crash bug scenarios, as well as the diverse prompt formulations and interaction strategies used.
This indicates that the model is not simply memorizing the data used in our study, even if the LLM had been exposed to this data. 
\rev{Furthermore, the effectiveness on previously unseen dataset demonstrates the experimental results in RQ1–RQ4 are not influenced by data contamination.}

\vspace{-1mm}
\section{Conclusion}
This study first empirically investigates the effectiveness of LLMs in resolving real-world environment-related crash bugs by comparing their performance with code-related crash bugs, and exploring the impact of different prompt strategies and multi-round interactions. 
Leveraging experimental findings, we introduce \app{}, a methodology that optimizes the interaction process and prompt design for resolving crash bugs. Our evaluation results demonstrate that \app{} achieves significant improvements in resolving environment-related crash bugs across multiple LLMs through its optimized interaction process.




\bibliographystyle{ACM-Reference-Format}
\bibliography{ref}

\end{document}